\documentclass[authoryear,preprint,review,12pt]{elsarticle_arxiv}

\usepackage{amssymb,amsmath}
\usepackage{mathabx}
\journal{Icarus}

\begin{document}

\begin{frontmatter}

  \title{Assessment of the 2880 impact threat from asteroid (29075)
    1950 DA}

  \author[jpl]{D. Farnocchia}
  \ead{Davide.Farnocchia@jpl.nasa.gov}
  \author[jpl]{S.~R. Chesley}

  \address[jpl]{Jet Propulsion Laboratory, California Institute of
    Technology, 4800 Oak Grove Drive, Pasadena, CA 91109, USA}

\begin{abstract}
  In this paper we perform an assessment of the 2880 Earth impact risk
  for asteroid (29075) 1950 DA. To obtain reliable predictions we
  analyze the contribution of the observational dataset and the
  astrometric treatment, the numerical error in the long-term
  integration, and the different accelerations acting on the
  asteroid. The main source of uncertainty is the Yarkovsky effect,
  which we statistically model starting from 1950 DA's available
  physical characterization, astrometry, and dynamical
  properties. Before the realease of 2012 radar data, this modeling
  suggests that 1950 DA has 99\% likelihood of being a retrograde
  rotator. By using a 7-dimensional Monte Carlo sampling we map 1950
  DA's uncertainty region to the 2880 close approach $b$-plane and
  find a $5\times10^{-4}$ impact probability. With the recently
  released 2012 radar observations, the direct rotation is definetly
  ruled out and the impact probability decreases to $2.5\times10^{-4}$.
\end{abstract}


\end{frontmatter}

\section{Introduction}
Near Earth asteroid (29075) 1950 DA was first discovered in 1950 by
C. A. Wirtanen at Lick Observatory \citep{discovery} and then lost for
more than 50 yr. In December 2000 the asteroid was rediscovered at
Lowell Observatory-LONEOS \citep{rediscovery} as 2000 YK$_{66}$ and
subsequently recognized to be 1950 DA.

In 2001, 1950 DA experienced an Earth close approach at 0.05 au and
radar observations were obtained from Arecibo and Goldstone. These
radar observations significantly reduced the orbital uncertainty and
allowed long-term predictions. In particular, \citet{giorgini02}
showed that there is a non-negligible probability (upper bound 0.33\%)
for an Earth impact in March 2880. The occurrence of such an impact is
decisively driven by the Yarkovsky effect, a subtle nongravitational
perturbation arising from the anisotropic re-emission at thermal
wavelengths of absorbed solar radiation. This perturbation causes a
secular variation in semimajor axis resulting in a mean anomaly runoff
that accumulates quadratically with time \citep{vok_2000}. As 1950 DA
experiences several planetary encounters
\citep[][Table~1]{giorgini02}, the runoff caused by the Yarkovsky
effect is amplified and therefore becomes important for 1950 DA's
predictions.

\citet{busch} use the 2001 radar observations to constrain the
physical properties of 1950 DA. The Ondrejov Asteroid Photometry
Project\footnote{http://www.asu.cas.cz/$\sim$ppravec/neo.html}
provides additional information on 1950 DA's physical model from
lightcurve observations obtained during the 2001 close approach.
However, the known physical characterization does not yet allow an
estimate of the Yarkovsky effect. In particular, the pole orientation
is still unknown and so is the sign of 1950 DA's orbital drift.

Because of the decisive contribution of the Yarkovsky effect, 1950 DA
belongs to a class of ``special objects'', which also includes
asteroids (99942) Apophis \citep{farnocchia_apophis} and (101955)
Bennu \citep{rq36}.  Each of these objects presents unique features
and demanding tasks. In particular, for 1950 DA we are pushing the
impact prediction horizon for a time interval that is four times
longer than ever analyzed for any other asteroid. Therefore,
performing the impact hazard assessment requires a specific effort and
the development of ad hoc techniques beyond what is routinely done by
the automatic impact monitoring systems
Sentry\footnote{http://neo.jpl.nasa.gov/risk} and
NEODyS\footnote{http://newton.dm.unipi.it/neodys}
\citep{neodys_sentry}.

\section{Orbital solution}
1950 DA has a long observed arc that allows a precise
estimate of the orbit. The earliest 18 observations
are from 1950. Then, we have two isolated observations in 1981 and
more than 450 observations from 2000 to 2012.  Moreover, in March 2001
the Arecibo and Goldstone observatories obtained 13 radar
observations\footnote{http://ssd.jpl.nasa.gov/?radar},
specifically 8 delay and 5 Doppler measurements \citep{giorgini02}.
(The contribution of the recently released 2012 radar observations is
discussed in Sec.~\ref{s:new_rad}.)

To properly handle the observation dataset and mitigate the effect of
star catalog systematic errors, we applied the debiasing and weighting
described by \citet{cbm10}, which we refer to as CBM10. Furthermore,
for observatories with $N>7$ observations on the same night we relaxed
the weights by a factor $\sqrt{N/5}$. This relaxation factor reduces
the contribution of batches containing a large number of observations,
e.g., Ondrejov lightcurve observations in late February 2001. Among
the post-2000 observations, there are batches showing unusually high
astrometric biases and therefore we removed all the batches with
apparent bias larger than 1". The discovery observation has a low
number of significant digits, so it was weighted at 30". Finally, we
applied weights at 2" to observations marked with the MPC flag `A',
i.e., when right ascension and declination in the J2000 system were
obtained by rotating the B1950 coordinates. Table~\ref{t:nom_sol}
contains the orbital elements corresponding to this astrometric
treatment, which is referred to as ``Nominal'' throughout the paper.
It is worth pointing out that this solution was computed without
accounting for the Yarkovsky effect, which is discussed in
Sec.\ref{s:nongrav}

\begin{table}
\begin{center}
\begin{tabular}{lcc}
\hline
Epoch & 2012 Sep 30.0 & TDB\\ 
Eccentricity & 0.5082852298(358) & \\
Perihelion distance & 0.8350375895(606) & au\\
Perihelion time & 2012 May 8.94652197(622) & TDB\\
Longitude of ascending node & 356.72810476(900) & deg\\
Argument of perihelion & 224.61346319(964) & deg\\
Inclination & 12.17480729(584) & deg\\
\hline
\end{tabular}
\end{center}
\caption{Orbital solution for asteroid 1950 DA. Numbers in
    parentheses indicate the 1$\sigma$ formal uncertainties for the
    corresponding digits in the parameter value.\label{t:nom_sol}}
\end{table}

Table~\ref{t:orb_sol} shows the normalized RMS according to different
observational datasets and astrometric schemes: a) observations only
from 1950 to March 2001, which is similar to the dataset used by
\citet{giorgini02} and which we refer to as the G02 dataset; b) G02
with the application of the \citet{cbm10} astrometric scheme (G02 + CBM10); c) the
full observational dataset with the application of the \citet{cbm10}
astrometric scheme (ALL + CBM10); d) the full observation dataset and the
astrometric treatment described above (Nominal). It is interesting to note that
the nominal solution provides the best match to the 2001 delay
measurements, which highlights the importance of using the full arc and
the goodness of the astrometric scheme described before. The table
also contains the coordinates on the 2880 $b$-plane, which is the
plane normal to the incoming asymptote of the geocentric hyperbola on
which the asteroid travels when it is closest to the planet
\citep{valsecchi}. The $\zeta$ axis is in the direction on the
$b$-plane opposite to the projection of the velocity of the Earth and
the $\xi$ axis completes the right-handed coordinate system. The G02
solution has the largest positive $\zeta_{2880}$, i.e., is the one
arriving at the close approach with the largest delay. The
\citet{cbm10} astrometric scheme and the use of the full arc
progressively decrease $\zeta_{2880}$, with the nominal solution
reaching the 2880 close approach with the largest advance with respect
to the Earth. Our nominal solution is 1.46$\sigma$ away from the G02
prediction, for which $\sigma_{\zeta_{2880}} = 3.19 \times 10^6$ km.

\begin{table}
\begin{center}
\begin{tabular}{lccccc}
  \hline
  Astrometry & NRMS & $\xi_{2880}$& $\sigma_{\xi_{2880}}$&
  $\zeta_{2880}$ & $\sigma_{\zeta_{2880}}$\\
   & in delay & [$10^3$ km] & [km] &  [$10^6$ km] &  [$10^6$ km]\\
  \hline
  G02 & 0.731 & 22.8 & 5.08 & 2.65 & 3.19\\
  G02 + CBM10 & 0.728 & 0.88 & 4.55 & 0.61 & 3.19\\
  All + CBM10 & 0.719 & 5.25 & 3.41 & -1.32 & 1.61\\
  Nominal & 0.706 & 11.8 & 3.56 & -2.01 & 1.65\\
  \hline
\end{tabular}
\end{center}
\caption{Normalized RMS (NRMS) for the 2001 radar delay measurements and $b$-plane
  coordinates for different observation datasets and astrometric
  schemes.\label{t:orb_sol}}
\end{table}

\begin{figure}
\centerline{\includegraphics[width=10cm]{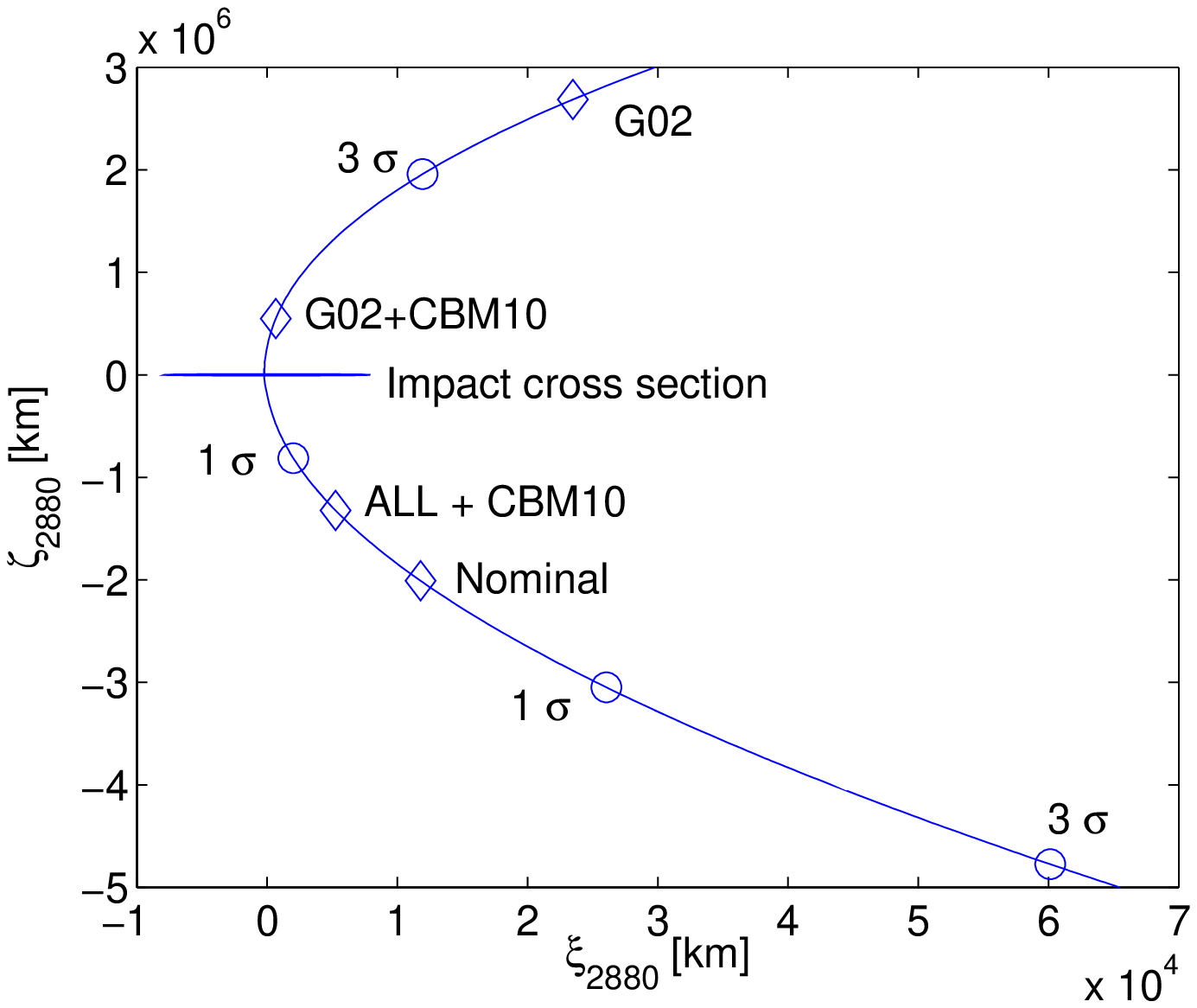}\includegraphics[width=10cm]{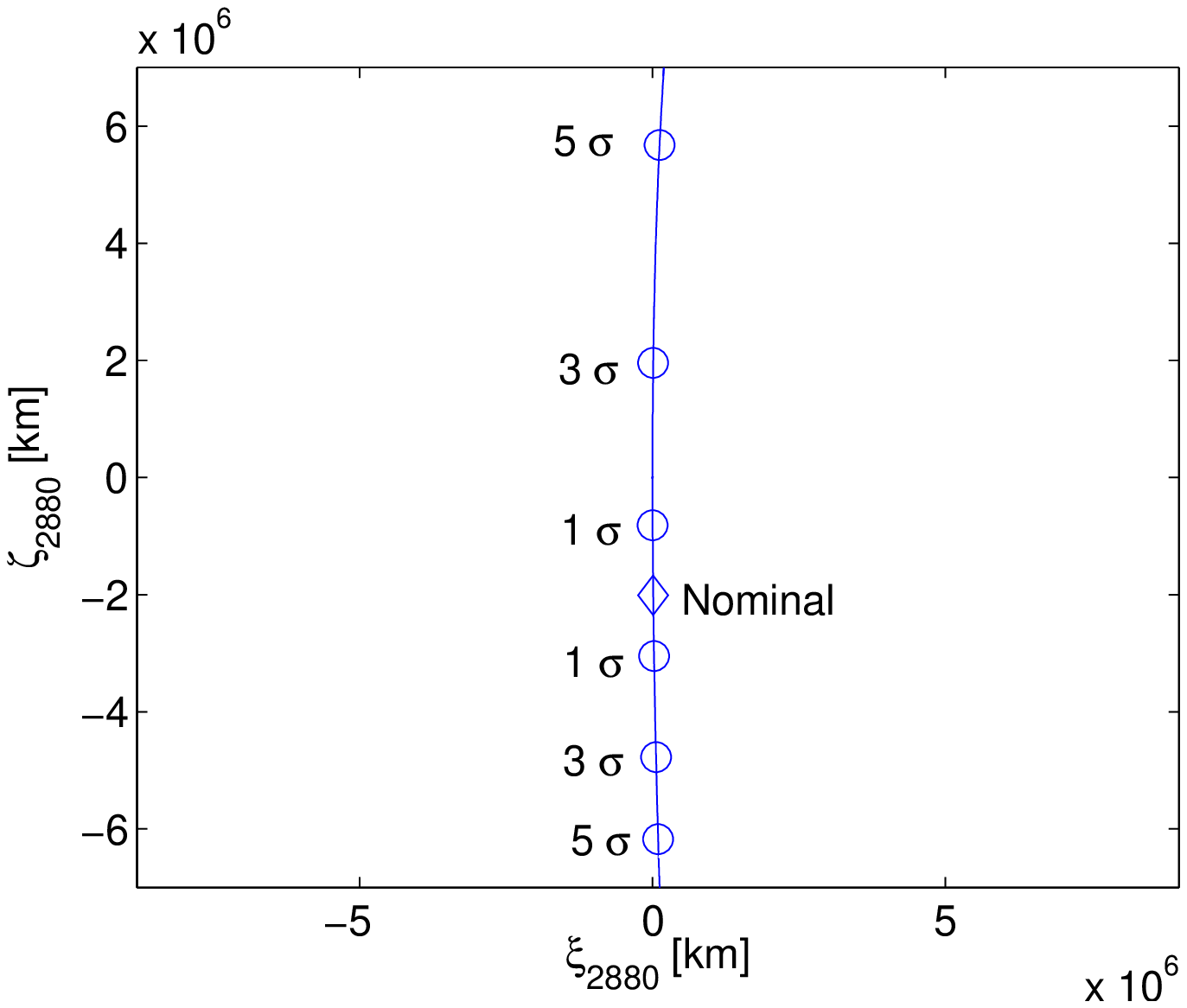}}
\caption{Line of variation in the 2880 $b$-plane. In the left panel
  diamonds are the $b$-plane coordinates of the orbital solutions
  corresponding to different observation datasets and astrometric
  schemes as described in Table~\ref{t:orb_sol}. Circles are 1, 3, and
  5$\sigma$ levels with respect to the nominal solution. The impact
  cross section has radius 1.24 Earth radii. The right panel shows
  the LOV and the $\sigma$ levels using the same scale for the
  axes.\label{f:lov}}
\end{figure}

Figure~\ref{f:lov} shows the uncertainty region on the 2880
$b$-plane. The 1$\sigma$ semi-width of the uncertainty region is 3--4
km, e.g., for $\zeta_{2880} = 0$ the full width is 7.12 km. Therefore,
we can perform the risk assessment by using a one dimensional
analysis. In particular, we consider the Line of Variation
\citep[LOV,][]{lov}, i.e., the line along which the uncertainty region
is most stretched and is therefore representative of the orbital
uncertainty. The nonlinearity of the mapping from the orbital
uncertainty space to the $b$-plane is evident from the curvature of
the uncertainty region and the locations of the $\sigma$ levels.
The Earth is at 1.64$\sigma$ from the nominal solution and a
simplistic computation of the corresponding impact probability (IP) is
5.06 $\times$ 10$^{-4}$. However, the computation of a reliable IP
requires a more careful analysis as discussed in the following
sections.

\section{Nongravitational perturbations}
\label{s:nongrav}
Nongravitational perturbations can play an important role for long
term predictions, therefore we used the following comet-like model
\citep{comet_nongrav}:
\begin{equation}
\mathbf a_{NG} = \left(\frac{r_0}{r}\right)^2 \left(A_1 \hat{\mathbf r} +
A_2 \hat{\mathbf t}\right)
\end{equation}
where $r_0 = 1$ au, $r$ is the heliocentric distance of the asteroid,
$\hat{\mathbf r}$ is the radial direction, and $\hat{\mathbf t}$ is
the transverse direction.

The radial component of $\mathbf a_{NG}$ models direct and reflected
solar radiation pressure and $A_1$ can be related to the asteroid's
physical quantities as follows \citep{vok_srp}:
\begin{equation}\label{e:A1}
A_1 = \left(1 + \frac{4}{9}A\right) AMR \frac{G_S}{c}
\end{equation}
where $A$ is the Bond albedo, $AMR$ is the asteroid's area-to-mass ratio, $G_S = 1371$ W/m$^2$
is the solar constant, and $c$ is the speed of light.

The transverse component of $\mathbf a_{NG}$ models the Yarkovsky
perturbation \citep{bottke_yarko} and $A_2$ can be related to the
asteroid's physical quantities as follows \citep{farnocchia_yarko}:
\begin{equation}\label{e:A2}
A_2 = \frac{4(1-A)}{9}\Phi(1 \mbox{au}) f(\Theta) \cos\gamma\; ,\ \
f(\Theta) = \frac{0.5\Theta}{1 + \Theta + 0.5\Theta^2}
\end{equation}
where $\Phi(1 \mbox{au})$ is the standard radiation force factor at 1
au, $\Theta$ is the thermal parameter, and $\gamma$ is the obliquity
\citep{vok_2000}. To drop the dependence of $\Theta$ on $r$, we computed
the subsolar temperature \citep{vok98} at the orbital semilatus rectum.

\subsection{Physical model}
\label{s:phys_mod}
Though $A_1$ and $A_2$ are unknown, we can generate a statistical
sample representing these two parameters starting from the available
information on 1950 DA's physical model.  Table~\ref{t:phys_par}
reports the known physical parameters from \citet{busch} and the
Ondrejov Asteroid Photometry Project. In particular, \citet{busch}
provide two models for 1950 DA's rotation state, i.e., direct and
retrograde.  As a result, the following analysis initially discusses
these two models separately, unifying them only in the next
subsection.  Note that the spin orientations found by \citet{busch}
correspond to an obliquity of 24.5$^\circ$ (for the direct model) and
167.7$^\circ$ (for the retrograde model). Therefore, the spin axis is
far from being in the orbital plane and the seasonal component of the
Yarkovsky effect can be neglected.

\begin{table}
\begin{center}
\begin{tabular}{lccc}
  \hline
  & Direct rotation & Retrograde rotation\\
  \hline
  Spin $(\lambda,\beta) \pm 5^\circ$ & $(88.6^\circ,77.7^\circ)$ & 
  $(187.4^\circ,-89.5^\circ)$\\
  Effective diameter $D$ $\pm$ 10\% & 1.16 km & 1.30 km\\
  Minimum bulk density $\rho_{\min}$ $\pm$ 10\% & 3.0 g/cm$^3$ & 3.5 g/cm$^3$\\
  Absolute magnitude $H$ & \multicolumn{2}{c}{17.55 $\pm$ 0.3}\\
  Slope parameter $G$ & \multicolumn{2}{c}{0.03 $\pm$ 0.1}\\  
  Rotation period $P_{rot}$& \multicolumn{2}{c}{2.12160 $\pm$ 0.00004 h}\\
  \hline
\end{tabular}
\end{center}
 \caption{Available physical characterization of 1950 DA. $\lambda$,
   $\beta$, $D$, and $\rho_{\min}$ are from \citet{busch}, $H$, $G$,
   and $P_{rot}$ from the Ondrejov Asteroid Photometry Project.\label{t:phys_par}}
\end{table}

\begin{figure}
  \centerline{\includegraphics[width=10cm]{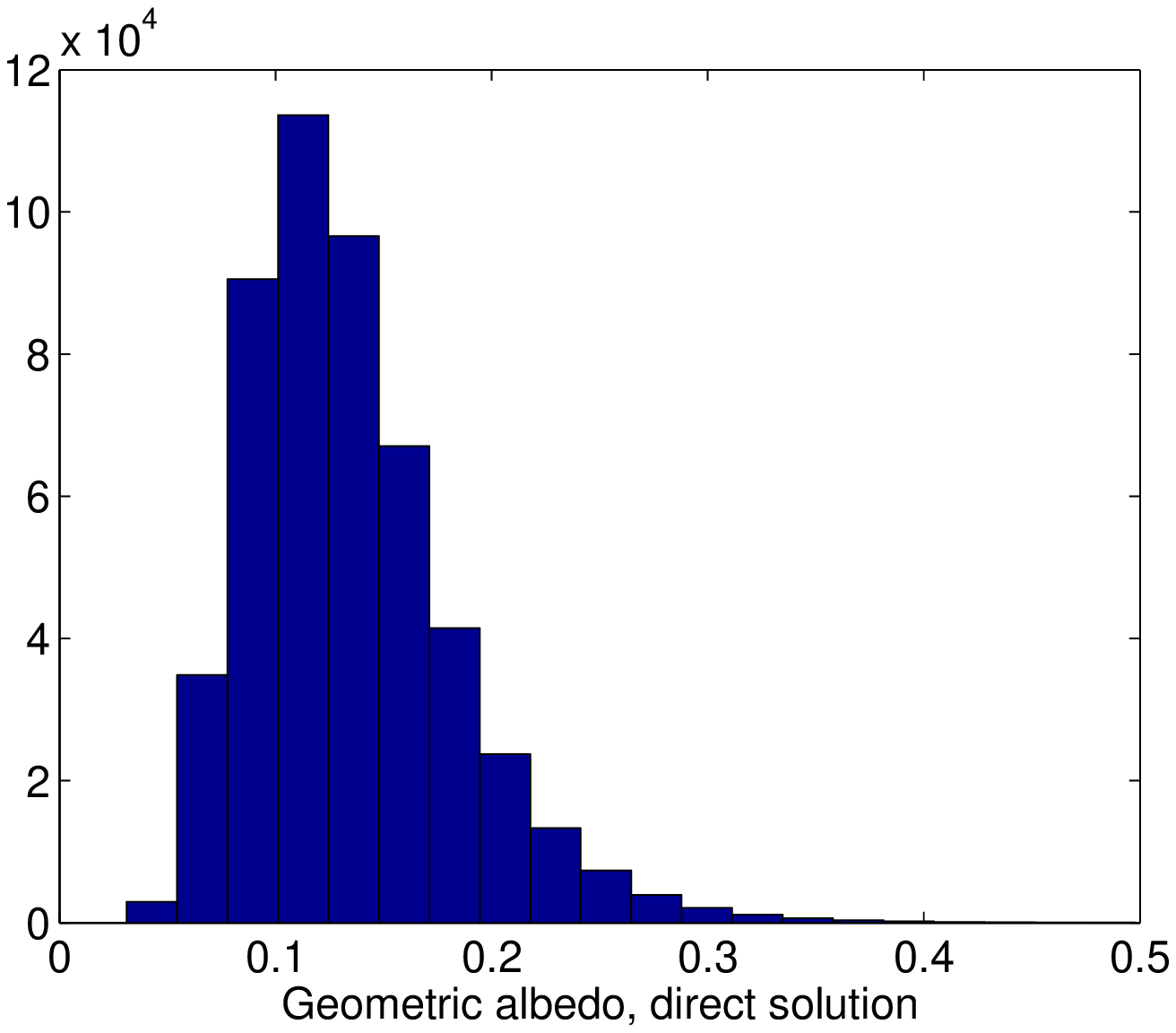}\includegraphics[width=10cm]{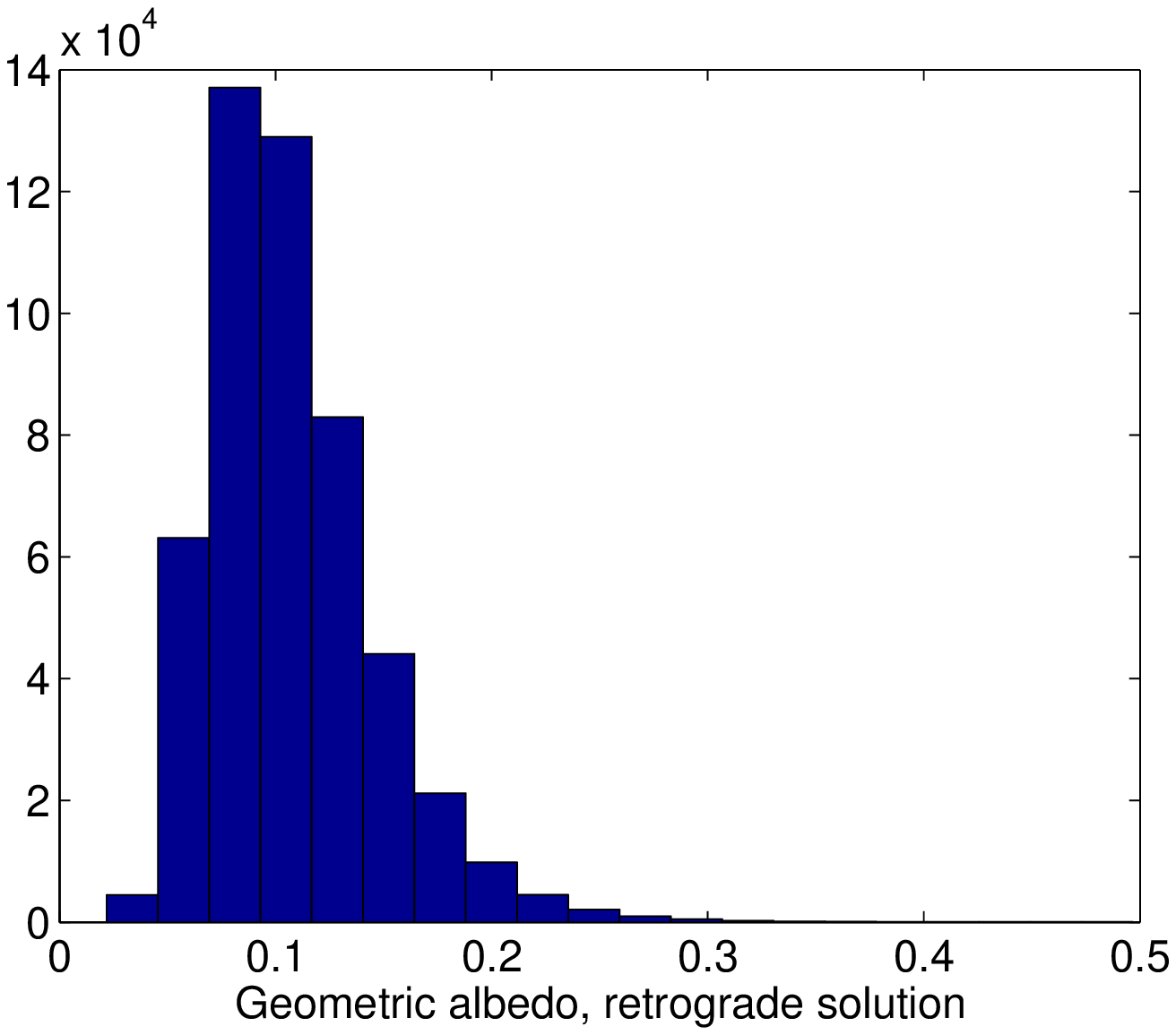}}
  \caption{Albedo distribution for both the direct and the retrograde
    models obtained by using the diameter from \citet{busch} and the
    absolute magnitude from the Ondrejov Asteroid Photometry
    Project.\label{f:albedo}}
\end{figure}

Figure~\ref{f:albedo} shows the geometric albedo $p_V$ distribution,
which was obtained from the absolute magnitude $H$ and equivalent
diameter $D$ according to \citet{pravec07}:
\begin{equation}
D = 1329\ \frac{10^{-0.2 H}}{\sqrt{p_V}}\, .
\end{equation}
The albedo distribution depicted in Fig.~\ref{f:albedo} is lower than
the geometric albedo reported by \citet{busch}, i.e., $p_V$ from 0.20
to 0.25. The reason for this difference is solely due to different
estimates of $H$. Indeed, \citet{busch} use photometry information
from 125 optical observations and obtain $H = 16.8$. However,
\citet{sloan} show the presence of biases in the known asteroid
absolute magnitudes catalog.  Therefore, we preferred to use the
absolute magnitude reported by the Ondrejov Asteroid Photometry
Project, i.e., $H=$17.55 $\pm$ 0.3. This value of $H$ also appears to
be more consistent with the 0.07 $\pm$ 0.02 geometric albedo reported
by NEOWISE \citep{wise}.

\citet{rivkin} suggest an EM taxonomic classification for 1950
DA. From the JPL Small Body
Database\footnote{http://ssd.jpl.nasa.gov/sbdb.cgi} we obtain that the
typical geometric albedo for E type asteroids is $\sim$ 0.4, therefore
the taxonomic type M (typical geometric albedo $\sim$ 0.17) seems more
likely.  By selecting M type asteroids from \citet{carry}, we obtain
an average bulk density $\rho = 3.86 \pm 0.87$ g/cm$^3$. Moreover,
\citet{busch} report a minimum bulk density (see
Table~\ref{t:phys_par}). Therefore, we used a truncated normal
distribution for $\rho$.

\citet{delbo} give a relationship between thermal inertia
$\Gamma$ and diameter $D$ (in km):
\[
\Gamma = d_0 D^{-\psi}\ ,  d_0 = ( 300 \pm 45) \mbox{ J m$^{-2}$ s$^{-0.5}$
  K$^{-1}$}\ ,\ \psi = 0.36 \pm 0.09\ .
\]
A preliminary thermal model of 1950 DA obtained from NEOWISE data
appears to be consistent with this relationship \citep{nugent}. From
$\Gamma$ and the $P_{rot}$ (see Table~\ref{t:phys_par}) we computed
the thermal parameter $\Theta$ according to \citet{vok98}.

\begin{figure}
  \centerline{\includegraphics[width=10cm]{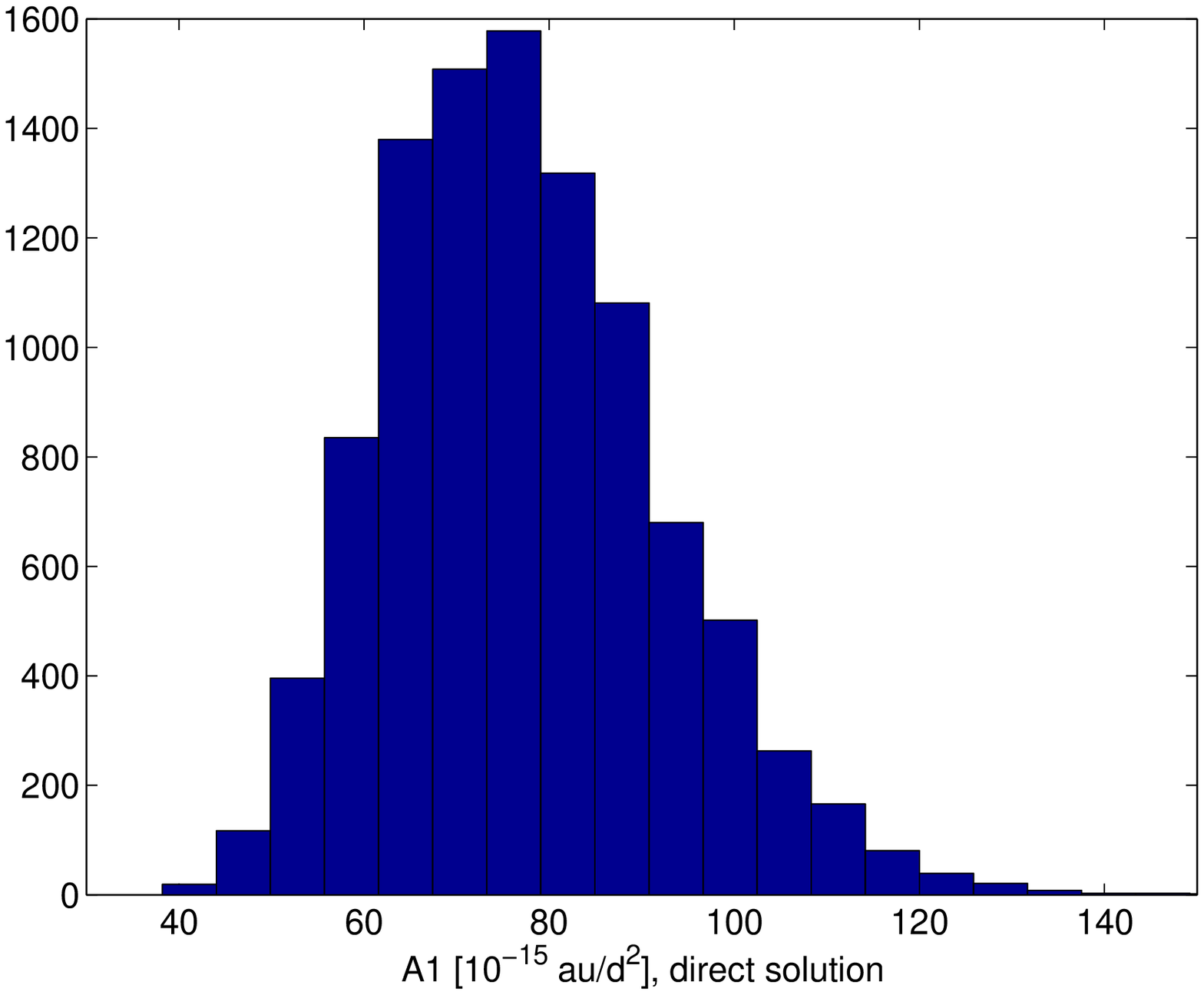}\includegraphics[width=10cm]{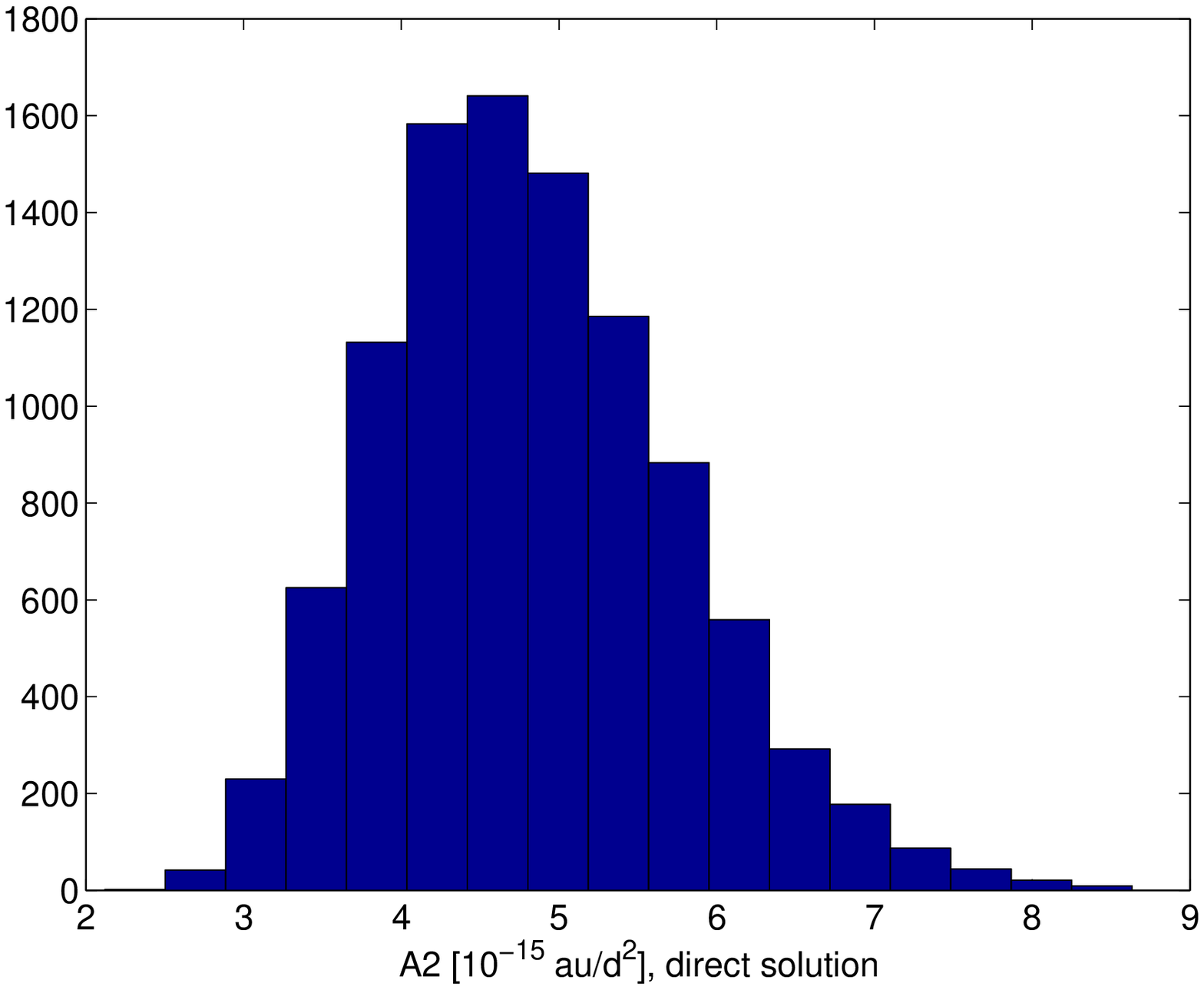}}
\caption{Nongravitational parameters sampling for the direct model.\label{f:A12_dir}}
\end{figure}

\begin{figure}
\centerline{\includegraphics[width=10cm]{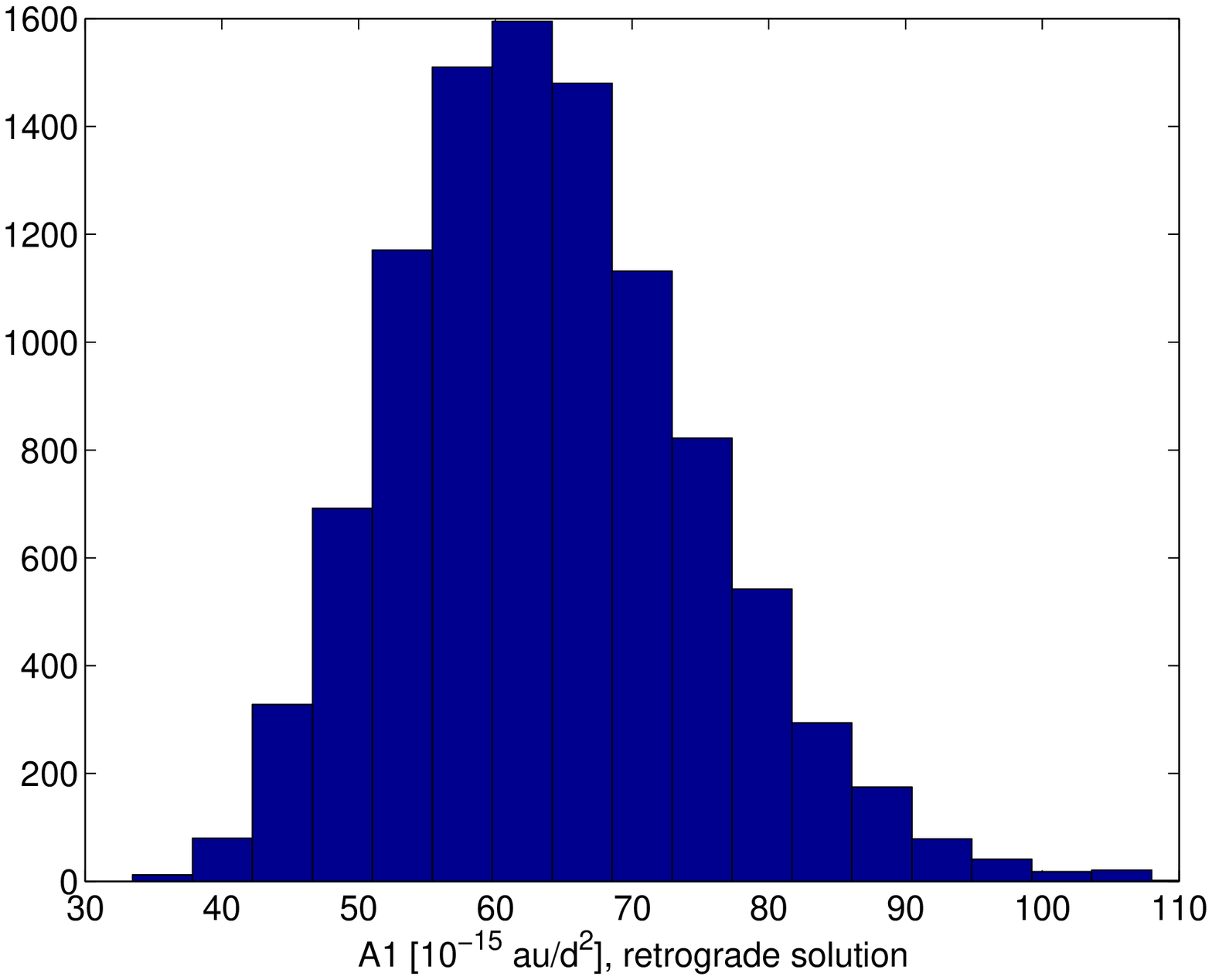}\includegraphics[width=10cm]{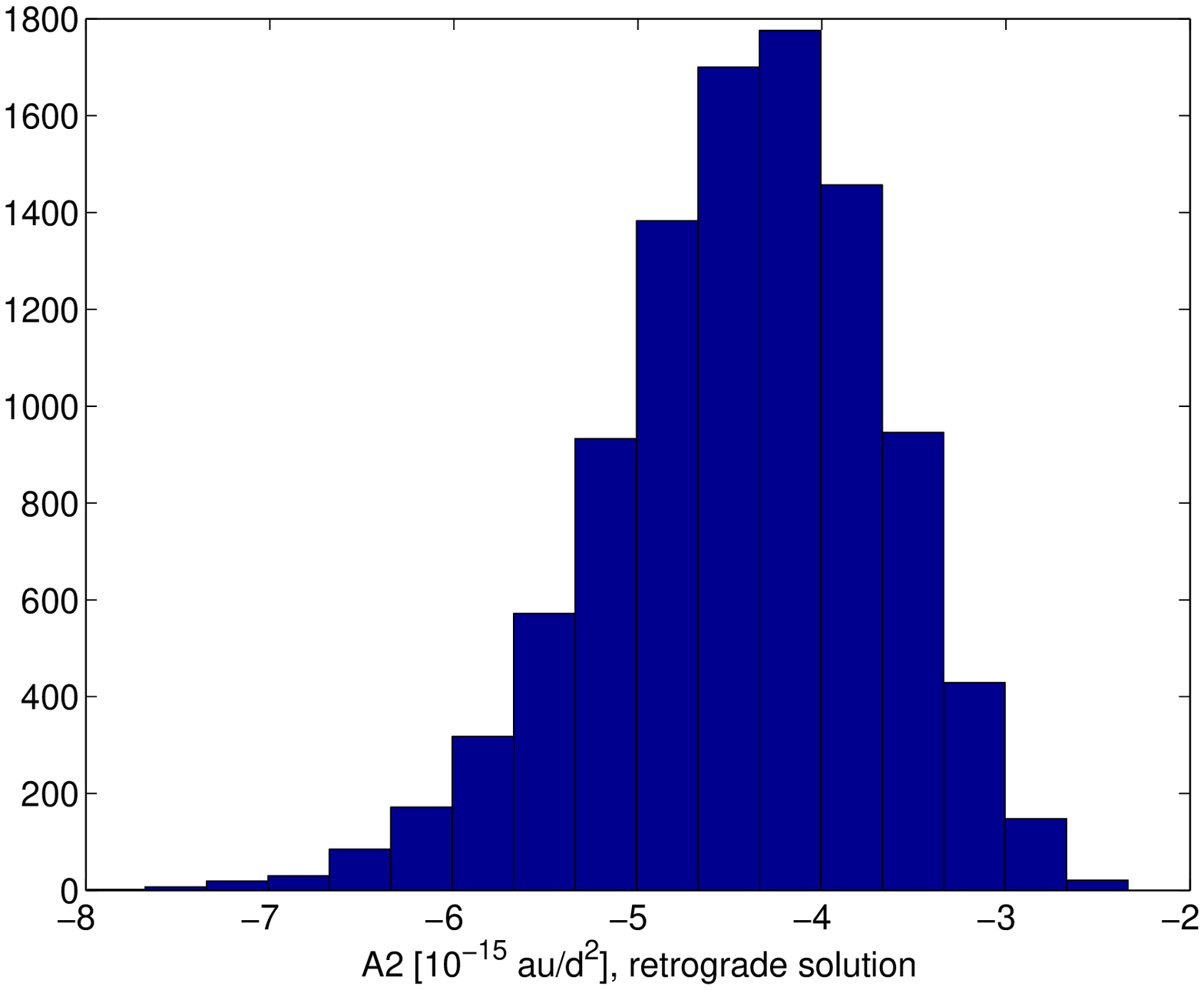}}
\caption{Nongravitational parameters sampling for the retrograde model.\label{f:A12_ret}}
\end{figure}

By using Eq.~\eqref{e:A1} and Eq.~\eqref{e:A2} we can map the physical
parameters described in this section and their uncertainties to the
nongravitational parameters $A_1$ and $A_2$. We then obtain the
distributions of Fig.~\ref{f:A12_dir} and Fig.~\ref{f:A12_ret} for the
direct and retrograde models.

\subsection{Overall distribution of the Yarkovsky effect}
\begin{figure}
\centerline{\includegraphics[width=10cm]{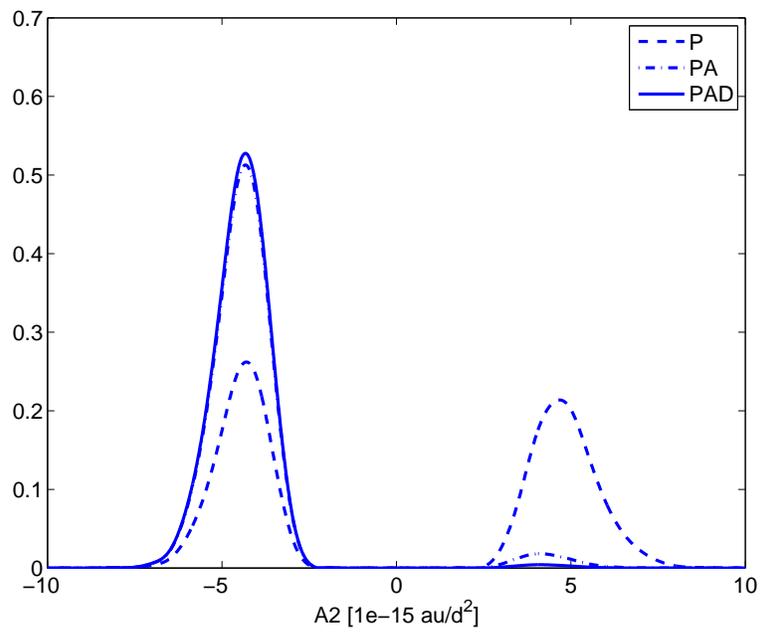}}
\caption{Distribution of $A_2$ obtained by using the physical model
  only (P), physical model plus astrometry (PA), and physical model
  plus astrometry plus dynamics (PAD).\label{f:A2}}
\end{figure}

The physical model provides us with two possible distributions for the
Yarkovsky parameter $A_2$. To obtain a single distribution of $A_2$ we
can use the following pieces of information from the following
sources:
%
\begin{itemize}
\item[(P)] According to \citet{busch}, both the direct and retrograde
  physical models provide good fits to radar and lightcurve
  data. Therefore, 1950 DA has a 50\% probability of being direct and
  a 50\% probability of being retrograde.
\item[(A)] The astrometry provides an additional constraint. In fact,
  if we solve for $A_2$ in the orbital fit to the observations
  \citep{farnocchia_yarko}, we find $A_2 = (-4.94 \pm 3.71)$ $\times$
  10$^{-15}$ au/d$^2$, which favors a retrograde rotation.
\item[(D)] The dynamical history of 1950 DA provides additional
  information. In fact, by using the \citet{population} Near Earth
  Object population model, we have that 1950 DA has a 63\% probability
  of coming from the $\nu_6$ resonance (Bottke, personal
  communication). As $\nu_6$ is at the inner edge of the Main Belt
  region, such objects can generally enter only by drifting inwards
  due to retrograde rotation. For the other NEO source regions we
  assume equal probability of entering by drifting inwards or outwards
  \citep{laspina}. Therefore, the probabilities that 1950 DA is direct
  or retrograde are 81.8\% and 18.2\%, respectively.
\end{itemize}
These are independent sources of information that can be used to
obtain the overall distribution of $A_2$. Figure~\ref{f:A2} shows how
the distribution of $A_2$ changes when the different pieces of
information are sequentially added. The distribution labeled with P
only uses the information from the physical model (i.e., 50\%-50\%
retrograde-direct ratio), PA uses also the astrometry. Finally, PAD
uses all the information above and is therefore the one we consider
most reliable.  Negative values of $A_2$ are predominant thus
suggesting that 1950 DA is likely to be a retrograde rotator with a
probability of 99\%. As a further confirmation, the geometric albedo
reported by NEOWISE is 0.07 $\pm$ 0.02, which somewhat favors the
retrograde solution (see Fig.~\ref{f:albedo}).

\section{Integration error}
\begin{figure}
\centerline{\includegraphics[width=10cm]{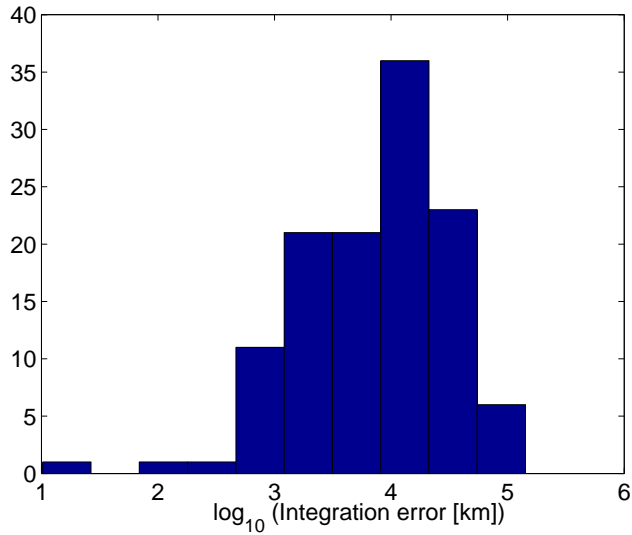}}
\caption{Distribution of the integration error from 2012 to 2880 for 121 virtual orbits
  along the LOV. The mean is $\sim$ 15000 km.\label{f:int_err}}
\end{figure}

The numerical integration produces a numerical error that can be
relevant for a long-term propagation. At each integration step we
introduce a random error below a fixed integration tolerance, which we
set to 10$^{-15}$. To estimate the numerical error in the propagation
through 2880 we compared the integration in double precision (which is
our default) to that in quadruple precision, assumed as the truth.
We made this comparison for 121 virtual asteroids (VAs) on the LOV:
each VA was propagated from the orbital solution epoch to 2880 in both
double and quadruple precision. Figure~\ref{f:int_err} shows the
distribution of the integration error in the logarithmic scale. The
integration error has a mean of $\sim$ 15000 km, but can be as large
as $\sim$ 150000 km.

\section{Dynamical error budget}
The long-term propagation through 2880 requires an assessment of the
relevance of the various perturbations affecting the dynamics of 1950
DA. Our dynamical model included:
\begin{itemize}
\item the Newtonian attraction of the Sun, eight planets, Pluto, and
  the Moon based on JPL's DE424 planetary ephemerides \citep{de424};
\item the Einstein-Infeld-Hoffman (EIH) relativistic approximation
  \citep{moyer} for the Sun, the planets, and the Moon;
\item the second order harmonics of the Earth gravity field for
  geocentric distance $<$ 0.01 au;
\item the Newtonian attraction of the 16 most massive asteroids
  ``BIG-16'' \citep[e.g., see Table~1 in][]{farnocchia_apophis}.
\end{itemize}

\begin{figure}
\centerline{\includegraphics[width=10cm]{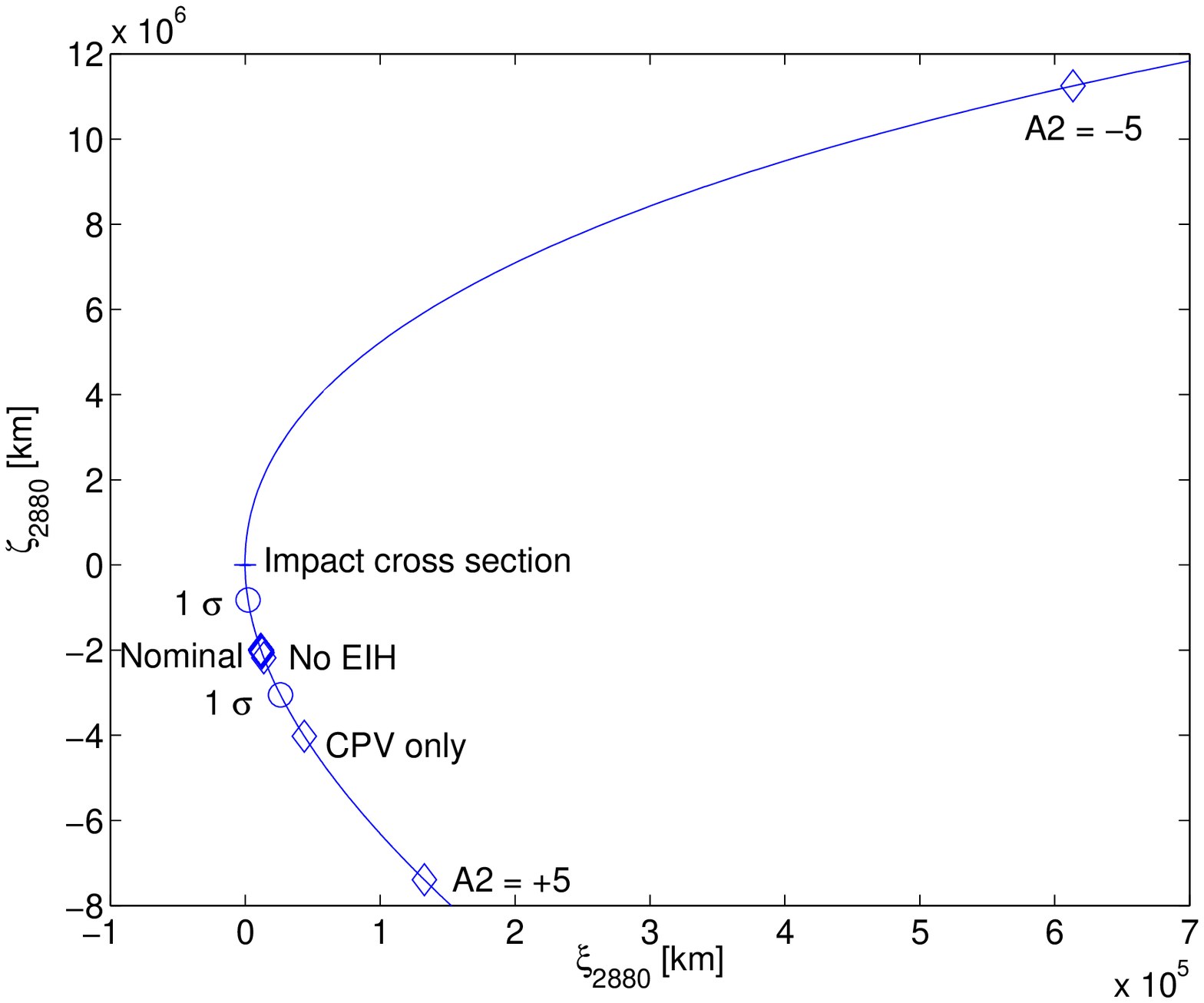}}
\caption{Coordinates on the $b$-plane for different settings of the
  dynamical model.\label{f:dyn_budget}}
\end{figure}

Figure~\ref{f:dyn_budget} and Table~\ref{t:dyn_budget} show the shift
in the 2880 $b$-plane coordinates for different settings of the
dynamical model. For each setting, we computed a corresponding
best-fitting orbital solution and propagated through 2880 to obtain
the $b$-plane shift with respect to the nominal prediction, which
corresponds to $A_1 = 0$ au/d$^2$ and $A_2 = 0$ au/d$^2$.

\begin{table}
\begin{center}
\begin{tabular}{lcc}
  \hline
  & $\Delta\xi_{2880}$ [km] & $\Delta\zeta_{2880}$ [$10^4$ km]\\
  \hline
  Sun only relativity (No EIH) & 2030 & -17.4\\
  Earth obl. (cut-off at 1 au) & 34.3 & -0.30\\
  CPV only & 32100 & -202\\
  BIG-16 + (78) Diana & 131 & 1.17\\
  BIG-16 + 9 pert. Bennu & 704 & -6.20\\
  $m_\Mercury$ +(-) 1$\sigma$ & -19.6 (159) & 0.17 (-1.41)\\
  $m_\Venus$ +(-) 1$\sigma$ & 161 (-177) & -1.43 (1.58)\\
  $m_\Earth$ +(-) 1$\sigma$ & -30.7 (-199) & 0.27 (1.78)\\
  $m_\Mars$ +(-) 1$\sigma$ & 506 (309) & -4.46 (-2.74)\\
  $m_\Jupiter$ +(-) 1$\sigma$ & -36.2 (-212) & 0.32 (1.90)\\
  $m_\Saturn$ +(-) 1$\sigma$ & -174 (-170) & 1.55 (1.52)\\
  $m_\Uranus$ +(-) 1$\sigma$ & -77.9 (133) & 0.69 (-1.18)\\
  $m_\Neptune$ +(-) 1$\sigma$ & 203 (-156) & -1.80 (1.39)\\
  $m_\Pluto$ +(-) 1$\sigma$ & -201 (-136) & 1.80 (1.21)\\
  $m_\Moon$ +(-) 1$\sigma$ & -135 (-145) & 1.20 (1.30)\\
  $A_1$ = 7 $\times$ 10$^{-14}$ au/d$^2$ & -118 & 1.05\\
  $A_2$ = 5 $\times$ 10$^{-15}$ au/d$^2$ & 120846 & -537\\
  $A_2$ = -5 $\times$ 10$^{-15}$ au/d$^2$ & 601720 & 1327\\
  \hline
\end{tabular}
\end{center}
\caption{Displacement of target plane coordinates for different dynamical settings.\label{t:dyn_budget}}
\end{table}

Among gravitational perturbations, the use of Ceres, Pallas, and Vesta
only (CPV) as perturbing asteroid produces a very large error. On the
other hand other perturbers such as (78) Diana, indicated by
\citet{giorgini02} as the perturbing asteroid experiencing the closest
approach to 1950 DA, and the nine additional asteroids considered by
\citet{rq36} for Bennu have a smaller contribution. The 1$\sigma$
variation of planetary masses and the Moon is rather small and
dominated by the integration error. Using 0.01 au as a cut-off for
including the Earth oblateness effect is a good approximation. In
fact, increasing the cut-off to 1 au has a negligible effect. On the
other hand, the use of a Sun-only relativistic model produces a
significant shift with respect to the nominal prediction.

\begin{figure}
\centerline{\includegraphics[width=10cm]{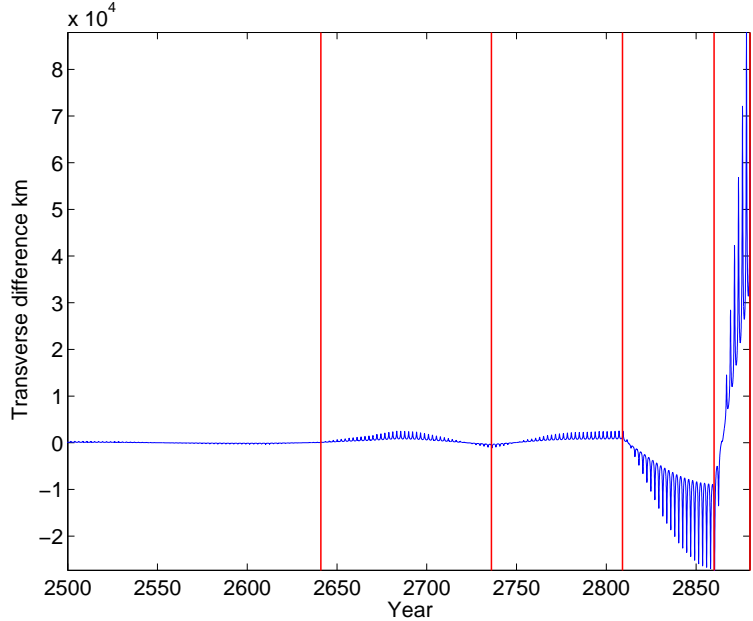}}
\caption{Along-track difference between a propagation with $A_2 = 5$
  $\times$ 10$^{-15}$ au/d$^2$ and a propagation without the Yarkovsky
  perturbation in the dynamical model. The vertical lines correspond
  to Earth close approaches with geocentric distance smaller than 0.1
  au.\label{f:yarko_trans}}
\end{figure}

For nongravitational perturbations, solar radiation pressure has a
negligible effect. This small effect can be explained by the fact that
the orbital fit to the observations corrects the semimajor axis to
compensate for the reduced gravitational parameter of the Sun
$GM'_\odot = GM_\odot(1-A_1r_0^2/GM_\odot)$, where $G$ is the
gravitational constant and $M_\odot$ is the mass of the Sun. Thus,
changing $A_1$ alters the semimajor axis but not the orbital
period. The Yarkovsky effect has the largest effect and is the main
source of uncertainty for the 2880 $b$-plane prediction. For
retrograde rotation, i.e., $A_2<0$, we have that $\zeta_{2880}$
increases. This behavior is counterintuitive, as a negative orbital
drift should imply a smaller period and thus an earlier arrival to the
2880 close approach. However, Fig.~\ref{f:yarko_trans} shows how Earth
approaches before 2880 can flip the uncertainty region and cause this
unexpected phenomena. It is important to note that to move the nominal
solution toward the Earth we need a retrograde rotation. Therefore,
the impact is much more likely with a retrograde rotation, which is
the opposite of the result obtained by \citet{giorgini02} and is due
to the different sign of $\zeta_{2880}$ for the two solutions (see
Fig.~\ref{f:lov})

We did not consider other perturbations, such as Galactic tide, Solar
mass loss, or Solar oblateness, as \citet{giorgini02} demonstrate that
the contribution of these perturbations is small and can be neglected.

\section{Risk assessment}\label{s:risk}
%
The impact risk assessment can be performed by means of a Monte Carlo
simulation. First, we randomly sampled $A_2$ according to the
distributions of Fig.~\ref{f:A2}. Then, for each value of $A_2$ we
computed the best fitting orbital solution by using our nominal
astrometric treatment and randomly selected a VA according to the
orbital covariance matrix. Finally, we propagated the VA
onto the 2880 $b$-plane.

\begin{figure}
  \centerline{\includegraphics[width=10cm]{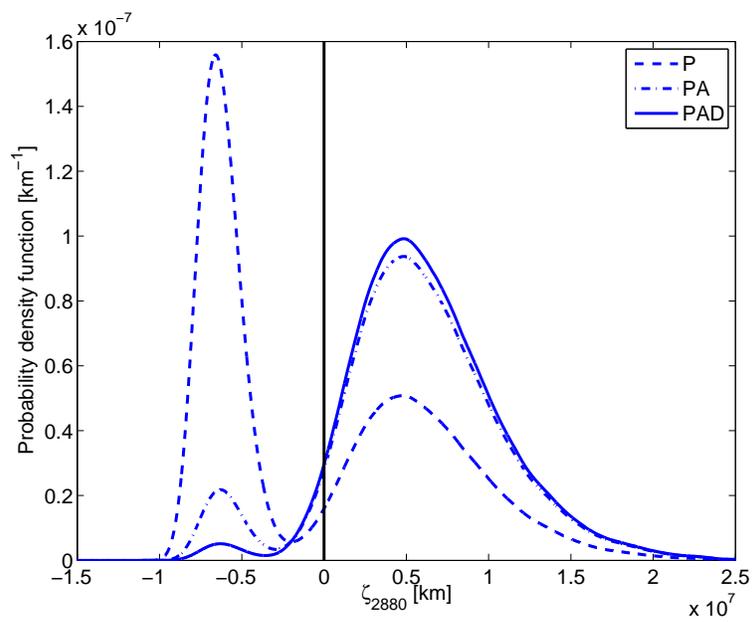}}
  \caption{Probability density function of $\zeta_{2880}$
    corresponding to the P, PA, and PAD distributions of
    Fig.~\ref{f:A2}. The vertical bar marks the location of the impact
    cross section of the Earth.\label{f:zeta}}
\end{figure}

Figure~\ref{f:zeta} shows the distribution of $\zeta_{2880}$
corresponding to the P, PA, and PAD distributions of Fig.~\ref{f:A2}.
The peaks on the left are related to the direct solution for the
rotation state and therefore the height decreases when more
constraints on $A_2$ are added and the retrograde solution becomes
more likely. On the other hand the peaks on the right correspond to
the retrograde solution.

The probability of an impact (IP) can be computed by multiplying the
$\zeta_{2880}$ probability density function (PDF) and the width $w$ of
the intersection between the LOV and the impact cross section. Note
that $w = 15856$ km, which is somewhat smaller than the diameter of
the impact cross section as the LOV does not pass directly through the
center of the Earth ($\xi_{2880} = -234$ km for $\zeta_{2880} =
0$). The best estimate of the IP is 4.69 $\times$ 10$^{-4}$, which is
given by the PAD solution. The corresponding Palermo Scale
\citep{palermo} is $-0.56$.

As reported by Table~\ref{t:ip_sensitivity} the IP is not very
sensitive to the amount of information used to constrain $A_2$.
Furthermore, Table~\ref{t:ip_sensitivity} shows the dependence of the
IP on the assumptions on 1950 DA's physical model. The IP always has a
similar order of magnitude, thus giving robustness to our result.

\begin{table}
\begin{center}
\begin{tabular}{lc}
  \hline
  & IP [$10^{-4}$]\\
  \hline
  PAD & 4.69\\
  P & 2.53\\
  PA & 4.44\\
  $H = 16.8$ $\pm$ 0.3 as in \citet{busch} & 5.73\\
  $H = 17.1$ $\pm$ 0.3 from MPC & 5.20\\
  $G = 0.15 \pm 0.1$ & 4.95\\
  $D = 1.5$ km $\pm$ 10\% & 6.74\\
  $D = 0.8$ km $\pm$ 10\% & 1.20\\
  $d_0 = 200$  $\pm$ 45  J m$^{-2}$ s$^{-0.5}$ K$^{-1}$ & 1.64\\
  $d_0 = 400$   $\pm$ 45 J m$^{-2}$ s$^{-0.5}$ K$^{-1}$ & 9.11\\
  $\rho = 2.5 \pm 0.87$ g/cm$^3$ & 1.03\\
  $\rho = 4.5 \pm 0.87$ g/cm$^3$ & 5.91\\
  \hline
\end{tabular}
\end{center}
\caption{Impact probability for different constraints on the
  Yarkovsky effect and different assumptions on the physical
  parameters of 1950 DA defined in Sec.\ref{s:phys_mod}.\label{t:ip_sensitivity}}
\end{table}

\section{Conclusions}
We found a $5\times 10^{-4}$ probability for an Earth impact of
asteroid (29075) 1950 DA in March 2880. The corresponding Palermo
Scale is $-0.56$, which is the highest among known possible asteroid
impacts. The long-term propagation calls for a detailed analysis of
all the possible sources of error, such as the astrometric treatment,
the dynamical model, and the integration error. Due to the related
secular variation in semimajor axis, the Yarkovsky effect plays a
decisive role in the risk assessment.  Even though the Yarkovsky
perturbation can be modeled from the available physical
characterization of 1950 DA, there is ambiguity in the rotation state
and therefore in the sign of the Yarkovsky related orbital drift. To
deal with this problem we introduced two additional constraints
related to the fit to the astrometric data and the dynamical history
of 1950 DA. Both these constraints suggest that the retrograde
rotation is more likely, with an overall $\sim$99\% probability. We
combined these two new independent sources of information with the
physical model to enhance our knowledge of the Yarkovsky effect, which
was in turn used to compute the probability of an impact in 2880. The
exceptional effort required to assess the impact threat from 1950 DA
outlines the importance of the impact monitoring as part of the
near-Earth asteroids tracking and risk mitigation.  Future radar
opportunities such as the 2032 close approach should confirm the spin
orientation and better estimate the Yarkovsky effect, thus resulting
in an improved risk assessment.

\section{Update}\label{s:new_rad}
After the paper was accepted, the 2001 radar observations were
remeasured and two 2012 Arecibo range measurement were released
\citep{new_radar}.  With the new data the astrometry provides a much
stronger constraint to the Yarkovsky effect, i.e., $A_2 = (-6.70 \pm
1.29) \times 10^{-15}$ au/d$^2$, and confirms that 1950~DA is a
retrograde rotator.

\begin{figure}
\centerline{\includegraphics[width=10cm]{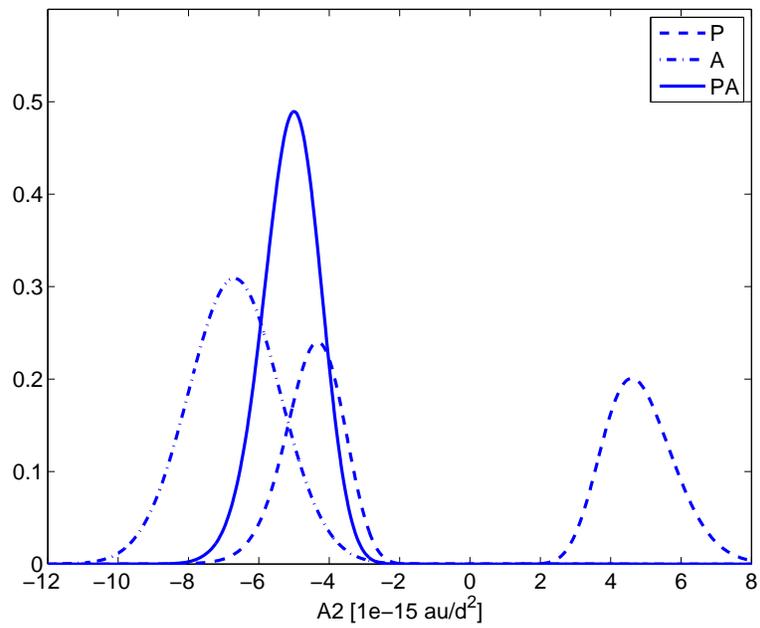}}
\caption{Updated distribution of $A_2$ obtained by using the physical model
  only (P), the astrometry only (A), and a combination of physical model
  and astrometry (PA).\label{f:A2_new}}
\end{figure}

Figure~\ref{f:A2_new} is an update of Fig.~\ref{f:A2}. We consider the
constraints from either the physical model (P) or the astrometry (A), as well
as the combination of them (PA).  The constraint from the dynamical
evolution of 1950~DA is not present as the astrometry already rules
out the direct rotation. By using the same technique described in
Sec.~\ref{s:risk} we can compute the corresponding impact
probabilities: $4.44 \times 10^{-4}$ for P, $5.05 \times 10^{-5}$ for
A, and $2.48 \times 10^{-4}$ for PA.

On one hand the astrometry suggests that the physical model is
underestimating the size of the Yarkovsky effect. A possible
explanation could be the presence of cohesive forces \citep{cohesion},
which would lower the minimum bulk density. Another possibility is
that the thermal inertia of 1950~DA is between 50 and 250 J m$^{-2}$
s$^{-0.5}$ K$^{-1}$, which is smaller than \citet{delbo} suggest but
would still make sense as the known NEA thermal inertias are quite
scattered.  On the other hand, the physical model suggests that $A_2$
is on the right side of the distribution obtained by the
astrometry. Therefore, we think that using both the astrometry and the
physical model (PA) still provides the most reliable solution. The
corresponding impact probability is $2.48 \times 10^{-4}$ and the
Palermo Scale -0.83.

\section*{Acknowledgments}
The authors thank A. Milani for his useful comments during the review
process.

DF was supported for this research by an appointment to the NASA
Postdoctoral Program at the Jet Propulsion Laboratory, California
Institute of Technology, administered by Oak Ridge Associated
Universities through a contract with NASA.

SC conducted this research at the Jet Propulsion Laboratory,
California Institute of Technology, under a contract with NASA.

Copyright 2013 California Institute of Technology. Government
sponsorship acknowledged.

\end{document}